\documentclass{egpubl}

 \JournalSubmission    
 \electronicVersion 

\ifpdf \usepackage[pdftex]{graphicx} \pdfcompresslevel=9
\else \usepackage[dvips]{graphicx} \fi

\PrintedOrElectronic

\usepackage{t1enc,dfadobe}

\usepackage{egweblnk}
\usepackage{cite}
\usepackage{verbatim}

\title{Deep Illumination: Approximating Dynamic Global Illumination \\with Generative Adversarial Networks}

\author[M. M. Thomas \& A. G. Forbes]
{\parbox{\textwidth}{\centering M. M. Thomas$^{1}$ and A. G. Forbes$^{2}$}
        \\
{\parbox{\textwidth}{\centering $^1$Electronic Visualization Laboratory, University of Illinois at Chicago, USA\\
         $^2$Computational Media Department, University of California, Santa Cruz, USA       }
}
}

\begin{document}

\teaser{
 \includegraphics[width=\linewidth]{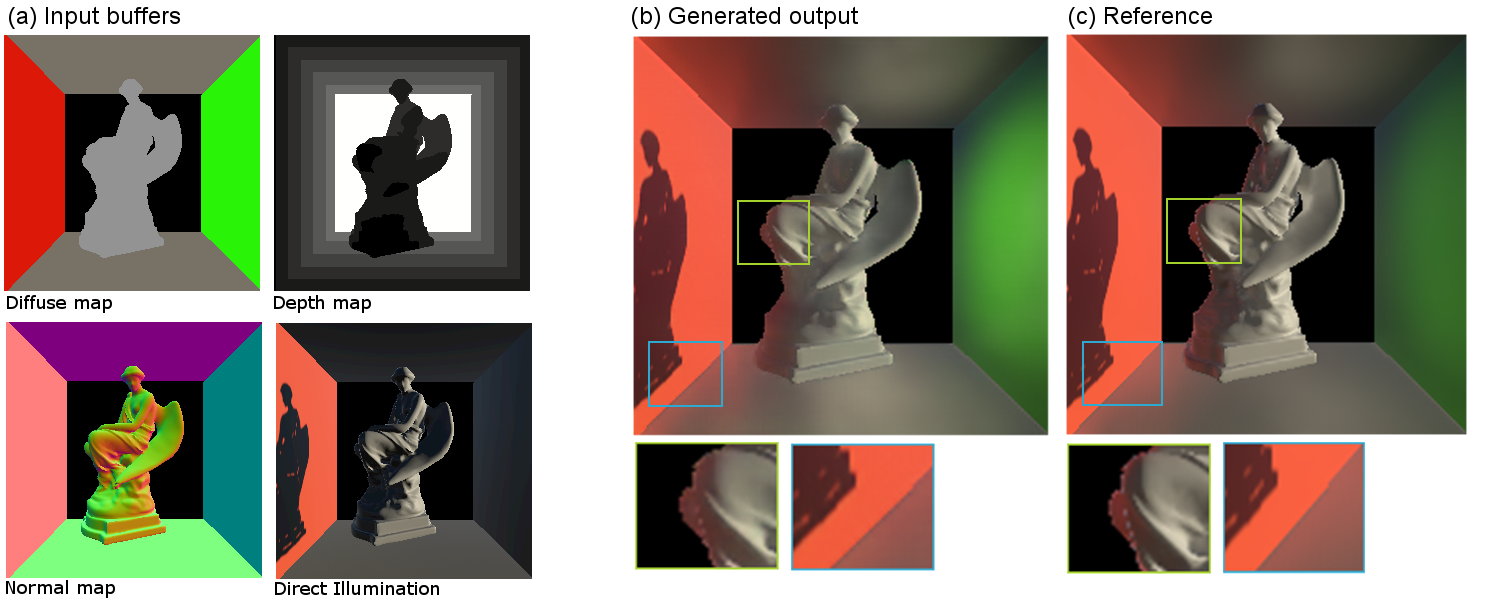}
 \centering
  \caption{The Deep Illumination technique produces global illumination output by training conditional generative adversarial networks (cGANs). Once the network is trained, inputting G-buffers along with the direct illumination buffer will add high-quality indirect illumination in real-time, including for novel configurations of a scene, such as new positions and orientations of lights and cameras, and entirely new objects.}
\label{fig:teaser}
}

\maketitle
\begin{abstract}

We present \textit{Deep Illumination}, a novel machine learning technique for approximating global illumination (GI) in real-time applications using a Conditional Generative Adversarial Network. Our primary focus is on generating indirect illumination and soft shadows with offline rendering quality at interactive rates. Inspired from recent advancement in image-to-image translation problems using deep generative convolutional networks, we introduce a variant of this network that learns a mapping from G-buffers (depth map, normal map, and diffuse map) and direct illumination to any global illumination solution. Our primary contribution is showing that a generative model can be used to learn a density estimation from screen space buffers to an advanced illumination model for a 3D environment. Once trained, our network can approximate global illumination for scene configurations it has never encountered before within the environment it was trained on. We evaluate Deep Illumination through a comparison with both a state of the art real-time GI technique (VXGI) and an offline rendering GI technique (path tracing). We show that our method produces effective GI approximations and is also computationally cheaper than existing GI techniques. Our technique has the potential to replace existing precomputed and screen-space techniques for producing global illumination effects in dynamic scenes with physically-based rendering quality.


\end{abstract}  
\section{Introduction}
Computing global illumination effects to reproduce real-world visual phenomena, such as indirect illumination~\cite{Ritschel2009, Nichols2010}, soft shadows~\cite{Scherzer2009,Eisemann2009,Ren2006}, reflections~\cite{Wald2002, Christensen2006}, crepuscular rays~\cite{engelhardt2010epipolar}, and caustics~\cite{Shah2007}, remains a challenge for real-time applications, even with the support of modern graphics hardware. Accurate physically-based global illumination computation is costly due to the difficulties in computing visibility between arbitrary points in 3D space and computing illumination at a point by taking the integration of light information from different directions~\cite{Ritschel2012}. In real-time applications, we are bounded by frame rates to maintain interactivity. Perceptually plausible illumination at real-time frame rates are often achieved by using approximations~\cite{Mara2016,Ritschel2009,Soler2010,Akenine-Moller:2008:RR:2829183}.

Global illumination (GI) approximation techniques can be classified as static or dynamic. In static approximation, light information is precomputed and stored as texture data, called lightmaps~\cite{Houska2006}. The advantage of using lightmaps is that they store the outputs of computationally expensive GI algorithms for determining light information, which are first computed offline and then readily available via fast texture lookups. The downside is that lightmaps are effective only for the static objects in a scene, for which the time-consuming computations only need to be done once. On the other hand, dynamic approximations support changes in lighting, soft shadows, and a range of other lighting effects. Screen space algorithms provide dynamic approximation, operating on G-buffers in image space rather than on 3D objects in model space. The G-buffers contain data about visible surfaces such as color, normal, depth, and position. The buffers are created during the first pass and then the actual shading is deferred to a second pass that uses only the buffer information~\cite{Saito1990}. Screen space algorithms are well suited for GPUs, which are designed for fast texture fetching. Although such algorithms can provide fast and dynamic GI solutions, image quality, level of detail, and accuracy may be compromised. 

Recent developments in deep learning and neural networks have been used to solve image-to-image translation problems using a generative model called Generative Adversarial Networks (GANs)~\cite{Goodfellow2014} and its conditional version (cGANs)~\cite{DBLP:journals/corr/MirzaO14,denton2015deep}. Both GANs and cGANs have shown promise in density estimation in an implicit fashion. So far, to the best of our knowledge, generative models have been used primarily in the domain of image processing, for example, to generate a plausible improvement in the resolution of an image~\cite{DBLP:journals/corr/LedigTHCATTWS16} , or to transform black-and-white images to color images~\cite{Isola2016}, or to automatically remove noise from image~\cite{DBLP:journals/corr/abs-1708-00961}. They have also been used to transfer features learned from a set of images (the ``style'') onto another image~\cite{DBLP:journals/corr/ZhangJL17,Gatys2016}. In a somewhat analogous manner, with this work we introduce the use of GANs to solve problems related to graphics rendering. We can view  effective GI approximation using image space buffers as a translation problem in which the network is expected to generate the GI output and where the image space buffers act as the input conditions. 

\begin{figure*}[htb]
  \centering
  \includegraphics[width=7in]{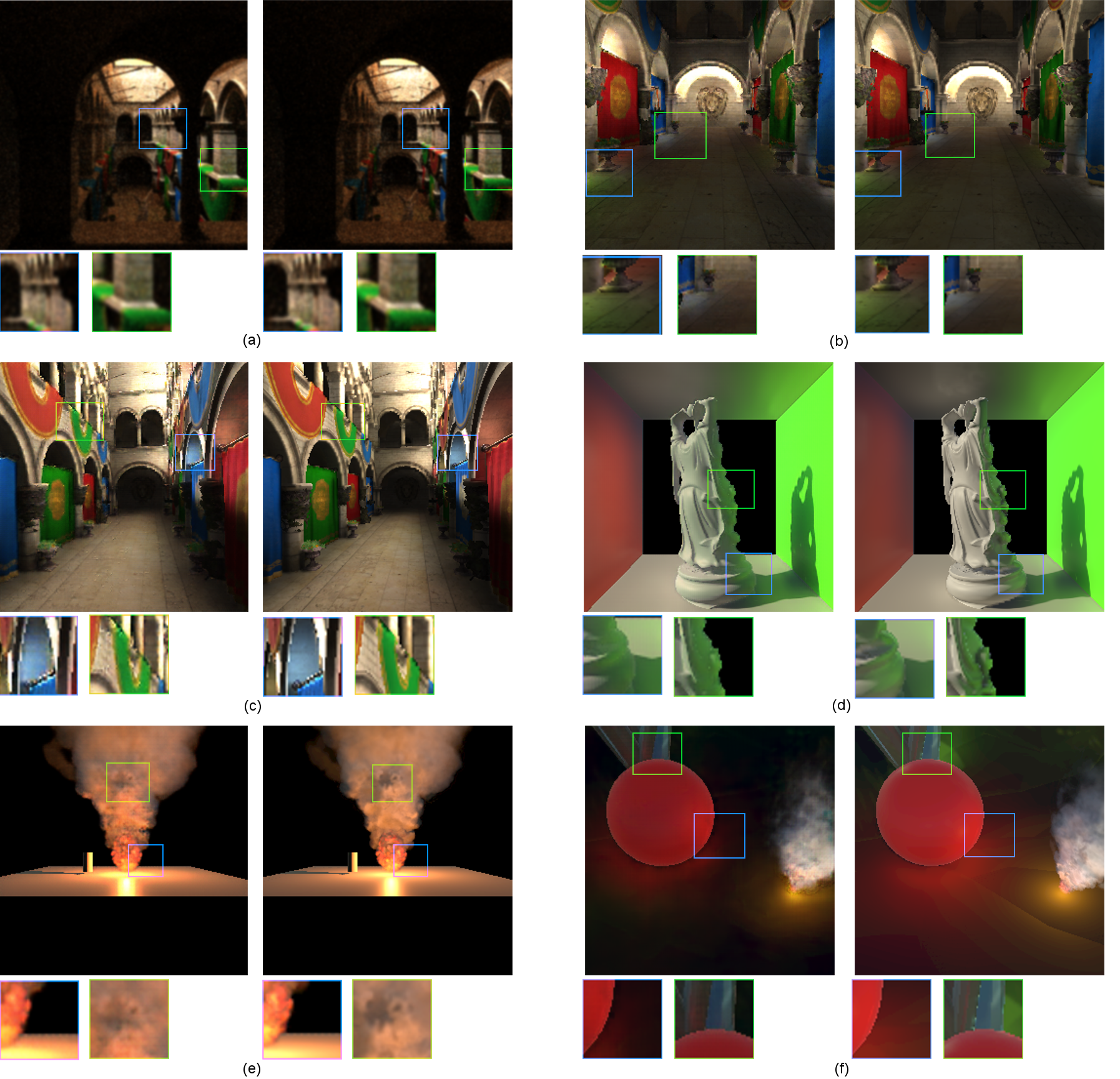}
  \caption{\label{figure:comparisons_and_closeups} 
  This figure shows the application of Deep Illumination to a wide number of real world scenarios, in which the subtleties of GI are captured nearly perfectly. In all examples, the left side is the Deep Illumination output, and the right side is the ground truth. We show close-ups to highlight comparisons showing interesting features in each scene. Pathtracing is used for example (a) and Voxel Cone Tracing is used for the rest of the examples. In (b), for example, we see that Deep Illumination can successfully generate the reflections of colored lights on the floor (blue square and green square). The example in (d) shows that our system can generate realistic global illumination ever for new objects that it was not trained on. 
  }
\end{figure*}

In this paper, we present a novel technique, \textit{Deep Illumination}, that computes indirect illuminations and soft shadows using a deep learning approach. The main idea is to train a network to learn high quality GI using an effective offline technique or online approximation technique. That is, similar to approaches for static approximation, we first precompute the lighting and make it available for retrieval when rendering frames in real time. However, we model this information in a neural network, where the network has learned a function to convert from one probability distribution to another. The key difference between our approach and a static approach, like generating lightmaps, is that we sample light information from a range of different possible configurations in terms of light direction, the position and orientation of the camera, and the position of objects in the scene we are interested in. We then extract the G-buffers and the direct illumination buffer for the same configuration and provide it as an input to a deep neural network in order to learn a mapping from these buffers to the computed GI. After training on a sufficient number of dynamic configurations, we show that our network has the ability to effectively approximate indirect illumination for object-light-camera configurations that it has not encountered during training. That is, it produces realistic GI even when the camera, lights, and objects in the scene are placed in new locations and/or have different orientations. It also provides strong temporal coherence in dynamic scenes so that the applied GI does not flicker from frame to frame. Not only does it produce effective GI for objects used during training in arbitrary positions, but good results are generated even for new objects that were not provided during training. Additionally, it learns to soften the hard shadows produced by direct illumination. \autoref{figure:comparisons_and_closeups} shows a range of different scenarios in which \textit{Deep Illumination} produces effective GI. In summary, the contribution of the paper is as follow:
\begin{itemize}
    \item We introduce \textit{Deep Illumination}, a novel technique for approximating GI in range of dynamic situations using a conditional generative adversarial network (Sec.~\ref{DeepIllumination});
\item We provide an evaluation of our technique versus state-of-the-art, in terms of both time and accuracy (Sec.~\ref{Evaluation});
\item We introduce the use of GANs for graphics rendering, and we provide a initial guidelines regarding the appropriate situations for which our approach and the use of adversarial networks is effective (Sec.~\ref{Discussion}). 
\end{itemize}

\noindent All code used in this research is open source and made freely available via our GitHub code repository, along with extensive examples and video documentation.

\begin{figure*}[htb]
  \centering
  \includegraphics[width=7in]{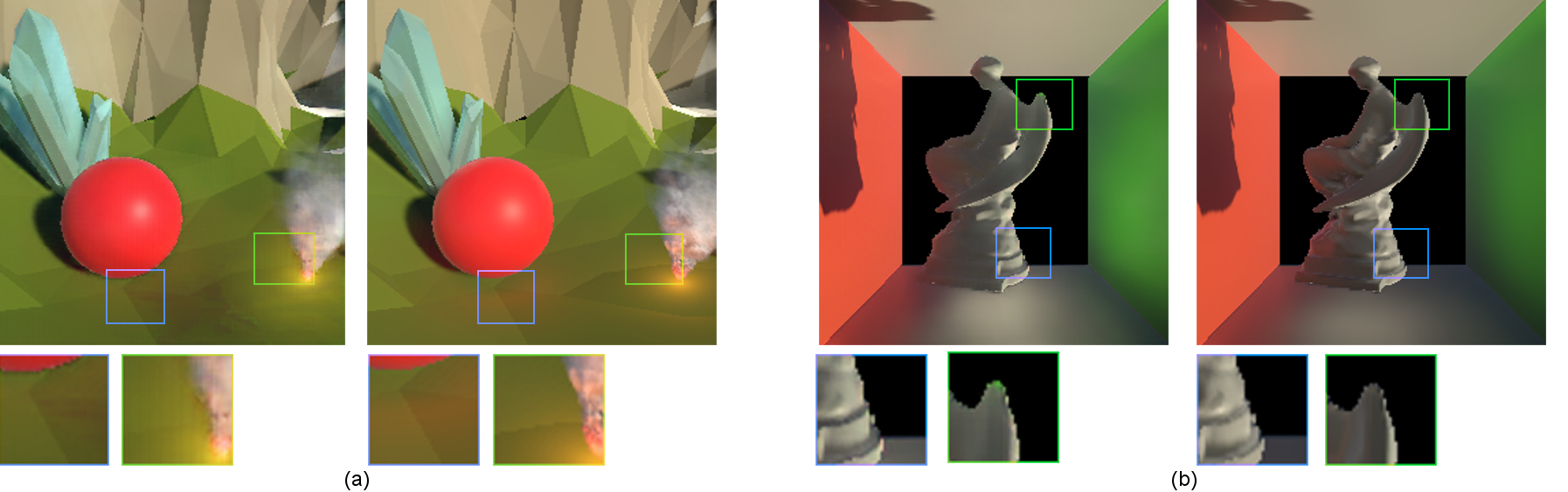}
  \caption{\label{figure:artifacts} This figure shows examples in which Deep Illumination generated rendering artifacts. As in \autoref{figure:comparisons_and_closeups}, the left side of both (a) and (b) is the Deep Illumination output, and the right side is the ground truth. In (a), we see that the crisp shadows (right side, blue square) are not reproduced properly (left side, blue square). Similarly, the particle system (green square) loses detail when compared to the ground truth. In (b), we see that our network has added realistic, but incorrect, indirect illumination on the base of the pedestal (blue square) an the tip of the angel's wing (green square). Note that, as in the example in \autoref{figure:comparisons_and_closeups}(d), our system was not trained on this object. Increasing the size of the training set or using more physically accurate ground truth techniques to train the network could minimize these artifacts (see Section 4). }
\end{figure*}

\section{Related Work}


Real-time global illumination is a perennial topic in computer graphics, and a range of approaches have been introduced to simulate real-world lighting effects. For example, precomputed radiance transfer~\cite{Sloan2002} makes use of precomputed light transport information with the assumption that the scene geometry is fixed and only light and camera changes. The main idea here is to store the light computation as spherical functions that is then projected into frequency space using spherical harmonics as a basis. The drawback of this approach is that we cannot change the position or shape of the object used for scene geometry without computing new coefficients of the spherical function. Also, spherical harmonics perform well with low-frequency features but tend to produce artifacts with high-frequency features. 

Another technique called Light Propagation Volumes (LPV) approximates single-bounce indirect illumination in large dynamic scenes~\cite{Kaplanyan2010}. Light information is stored in a 3D grid in the form of spherical harmonics. Light is iteratively propagated from each grid position to the adjacent ones where each grid represents the distribution of indirect light in the scene. Even though LPV is fast and stable, low resolution of the light propagation volume results in light bleeding, and it only allows for one light bounce. With our technique, as we discuss in Sec.~\ref{LOSSFUNCTION}, it is possible to train the network with ground truth generated from multiple light bounces and with high-frequency features without compromising performance or stability. 

Voxel cone tracing (VXGI) approximates one bounce of light in a dynamic scene~\cite{Crassin2014}. Here, instead of utilizing the actual scene geometry, a voxel representation is created once for the static geometry and then again per frame for dynamic objects. First, the scene geometry is converted to a map which encodes approximate opacity of space, averaged over voxels. This is followed by an emittance voxelization step, where emittance voxels store the amount of light the geometry emits into all directions. Finally, irradiance is calculated using cone tracing~\cite{Crassin2014}. While VXGI produces a good approximation of GI, cone tracing is an expensive step which adversely affects performance. As we describe in Sec.~\ref{SystemOverview}, we use VXGI as one of the techniques to generate ground truth for our network to learn. After training, our network can produce VXGI-quality output much faster than VXGI itself (see Sec.~\ref{NumberOfLayers}).

\subsection{Deep Learning Rendering Techniques}

Although, to our knowledge, \textit{Deep Illumination} is the first rendering technique that makes use of generative adversarial networks, deep learning approaches have recently been developed to address issues in computer graphics. One such approach is used to approximate ambient occlusion. \textit{NNAO}~\cite{Holden2016} is a fast, accurate screen space ambient occlusion technique that makes use of a neural network to learn an optimal approximation required to produce ambient occlusion effects. A four-layer neural network is used to learn a mapping from depth and camera space normal to an ambient occlusion value. Another technique called \textit{Deep Shading}~\cite{Nalbach2017} uses a convolutional neural network to learn a mapping from screen space buffers to various screen space effects, such as ambient occlusion, indirect light scattering, depth of field, motion blur, and anti-aliasing. Our technique differs from \textit{Deep Shading} in three important ways. First, while \textit{Deep Shading} uses a single convolutional neural network for training, we make use of two convolutional neural networks (a generator and a discriminator network), which has the advantage of producing crisp images via the incorporation of an adversarial loss function. Second, indirect illumination in \textit{Deep Shading} is trained with ground truth images computed in screen space; we show that our technique can either utilize ground truth images produced by any real-time or offline GI solution in order to produce high-quality output. Third, a \textit{Deep Shading} network is trained on a variety of data from different scenes, whereas we introduce a ``one network for one scene'' approach where we extract data from a single scene. This helps in preserving the style used for that scene during the approximation. Another advantage of this approach is that the network knows more information about one scene which is useful in making sound assumptions during the approximation, and which produces physically plausible outputs which are closer to the ground truth (See Sec.~\ref{Discussion} for a discussion of this approach).

\section{The  Technique}
\label{DeepIllumination}
In this section, we first describe how each stage of our \textit{Deep Illumination} technique works. We then explain how we extracted the data for training, along with a general guideline for using our technique on other datasets. We then describe the architecture of the neural networks used in \textit{Deep Illumination}, and finally we summarize the training phase, explaining the loss function used for learning global illumination features.


\subsection{System Overview} \label{SystemOverview}
We introduce a ``one network for one scene'' approach for training the network to learn GI features for a particular 3D scene. This approach is highly relevant for scenes that are dynamic, but which nonetheless have an clearly recognizable style with features that can be learnt by a neural network. For example, video games generally distinguish themselves with a distinct artistic style, and a different \textit{Deep Illumination} network could be created for each game. In video games where each level is artistically different from each other in terms of shading style and light setup, a \textit{Deep Illumination} network could be created for each level. 

There are two phases to our approach: 1) the training phase, and 2) the runtime phase. For the training phase, we extract G-buffers (depth map, normal map and diffuse map), the direct illumination buffer, and the output from an any expensive GI technique, which we use as our ``ground truth,'' from a 3D scene for different possible light-camera-object configurations in terms of position and direction. (We have used trained on the output of path tracing~\cite{Lafortune1993,Tokuyoshi2012} and Voxel-based Global Illumination~\cite{Crassin2014}, but our technique will work on any GI technique.) The GAN network is then trained in a semi-supervised fashion where the input to the network is the G-buffers and direct illumination buffer. The ground truth, that is, the output from the chosen GI technique, is used by the network to learn the distribution with the help of a loss function. Once training is complete, we can then use the trained network for the runtime phase. During the runtime phase, the network can generate indirect illuminations and soft shadows for new scene configurations at interactive rates, with a similar quality to the ground truth GI technique used for training, and using only image space buffer information. 

\subsection{Data Acquisition}
We prepared the dataset for two different scenarios (that is, we create a different GANs for each of the two scenes). In the first scenario, the scene has a fixed camera, and consisted of a Cornell box with a directional light rotating along an axis. A 3D model is placed in the center of the Cornell box, which rotates along multiple axes. For training and validation, we used only 3D models of a sphere, cylinder, cube, the Stanford bunny, and the Stanford dragon. We extracted 8000 pairs of G-buffers (depth, normal, and diffuse maps), the direct illumination buffer, and the ground truth from VXGI~\cite{Crassin2014} and GPU path tracing~\cite{Tokuyoshi2012}. All image pairs in the dataset have a resolution of 256x256 pixels. For testing, we extract 2000 image pairs with a statue model and the Happy Buddha model, both of which were not used in training. 

The second scenario presents a custom low-poly world, traversed with a moving camera and containing moving objects that are emissive in nature, along with particle effects. The light in the scene simulates the sun, producing a day-night cycle. (\autoref{figure:comparisons_and_closeups}f and \autoref{figure:artifacts}a show examples of this scenario.) We extracted 12,000 pairs of image space buffers. For training validation and testing, we used the same objects (unlike in the first scenario), but here the camera moves in a different path not encountered during training. The validation set is thus a mix of both unseen configurations of the scene and intermediate scenes from the training set. (Additionally, another portion of this scenario is used for evaluation purposes, as is explained Sec.~\ref{Evaluation}).

\begin{figure*}[htb]
  \centering
  \includegraphics[width=\linewidth]{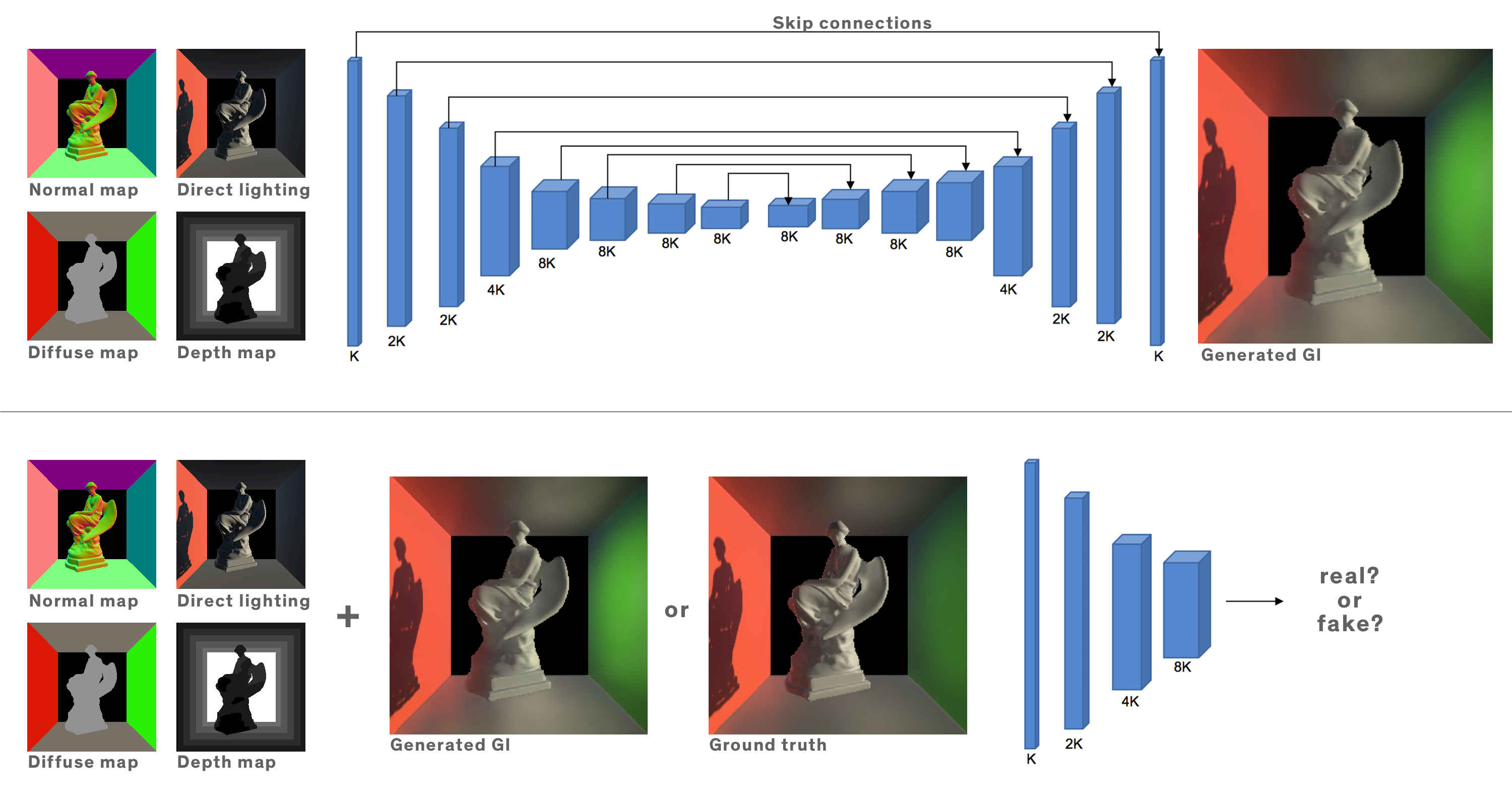}
  \caption{\label{fig:networks}
           This figure shows a schematic of the generator network (top) and discriminator network (bottom) used in Deep Illumination to learn a mapping from image space buffers to the target global illumination solution.}
\end{figure*}

\subsection{Network Architecture}
We use conditional generative adversarial networks (cGANs), which consists of a \textit{generator} and a \textit{discriminator}. The generator network takes G-buffers and the direct illumination buffer as the input and attempts to map it to its corresponding GI output. The discriminator network takes either the ground truth GI output or the generated GI output from the generator, and classifies them as real or fake, where ``real'' means that the input is from the distribution of images produced by the expensive GI technique. Thus the generator and discriminator network play a min-max adversarial game where the generator tries its best to fool the discriminator into thinking that the generated output is from the real distribution and the discriminator tried to learn from the real and generated image to classify them. This game will continue until discriminator is unable to distinguish between real and generated images~\cite{Goodfellow2014,DBLP:journals/corr/MirzaO14,Gauthier2014}.

\subsubsection{Structure of Generator Network} Our generator is a U-Net~\cite{Ronneberger2015}, which is a deep convolutional network consisting of an encoder and decoder with skip connections. The encoder consists of a convolution followed by an activation function, compressing the data into its high-level representation. The decoder does the opposite of the encoder, consisting of a deconvolution and an activation function. Both the encoder and the decoder undergo batch normalization, except for the first and last layers of the network. LeakyReLU~\cite{Maas2013} is used as the activation function of the encoder part, and ReLU and tanh (for the last layer only) are used in the decoder. Skip connections combine the output of an encoder to the input of the decoder counterpart. These skip connections are useful in recovering the spatial information lost during compression~\cite{He2016}. The input to the generator network is the G-buffers and direct illumination buffer concatenated into a 12 channel image, and the output is a 3-channel RGB image. We use 8 encoders and 8 decoders with a varying number of base layers (K=32, 64, 128) for evaluations. \autoref{fig:networks} (top) shows an overview of the generator network used in \textit{Deep Illumination}.

\subsubsection{Structure of Discriminator Network}
Our discriminator is a patchGAN which is also made up of deep convolutional layers, but patchGAN has the advantage that it gives a patch-wise probability of being real or fake, which is then averaged to get the output~\cite{Isola2016}. The discriminator consists of 5 encoders (no decoders) with LeakyReLU and sigmoid (only for the last layer) as the activation function. Similar to the generator network, the discriminator encoder undergoes batch normalization throughout, except for the first and last layers. The discriminator takes two sets of inputs, conditional image space buffers combined with the target image, or instead with the generated image. The discriminator then decides whether the given set is generated or not. The input to the discriminator is image space buffers and an image (real or generated) making it a 15 channel input, and output is a probability of how much of the input is real. We used a base layer k=64 for the discriminator. \autoref{fig:networks} (bottom) shows an overview of the discriminator network used in \textit{Deep Illumination}.

\subsubsection{Loss function} \label{LOSSFUNCTION} In addition to the original GAN loss function, we also use an L1 loss, which is a pixel-wise absolute value loss on the generated output. Unlike the traditional GANs, where the goal is to produce novel content as output, we care about the translation from image buffers to a final image rather than novel content generation. If we use GAN loss alone, we will end up with an image from the distribution we are looking for, but which lacks accuracy in terms of structure and light information~\cite{Isola2016}. This is because the discriminator is only designed to check whether the generated image is from the target probability distribution. If we use L1 loss alone, i.e., only a generator, then the network tends to produce blurry images, since it does well at low frequencies but fails at high frequencies that are needed to provide crispness, edges, and detail. The combination of these two loss function ensures that the network will produce crisp images that are from the desired distribution and that are accurate when compared with ground truth.

\subsection{Training}

Our network is implemented in PyTorch, and training is performed for 50 epochs, which takes approximately 3 hours on a Nvidia P5000 GPU. Random frames are selected from the dataset in mini-batches of 4 and the cost function is computed for each of these mini-batches. The weights of the network are updated using an adaptive gradient descent algorithm. We use a learning rate of 0.0002, and to avoid overfitting we use a dropout of 0.5 for the generator.

\section{Evaluation} \label{Evaluation}

To explore the capabilities of our approach, we tested our technique on a variety of network and dataset configurations. Our primary evaluation setup consists of the Crytek Sponza scene with a directional light acting as the sun. A camera moves around the scene in all possible directions extracting the screen space buffers (depth map, normal map, diffuse map and direct light) which will be the input to the network and output produced by VXGI, which will act as the ground truth (although in Sec.~\ref{DifferentGTs} we also describe a comparison of the learning convergence and accuracy of VXGI versus path tracing). The training and test set was carefully curated such that the training set will contain frames from point A and point B in the scene and the test set will contain all the intermediate frame from point A to point B. We evaluate the approximation based on the result generated from these unseen intermediate frames. Based on a survey of relevant literature and from our previous experience with neural network and generative model we generated the following hypotheses:

\begin{enumerate}

\item[\textbf{H1}] The number of training iterations will reduce mean square error as well as  perceptual error.
\item[\textbf{H2}] The neural network will produce a better approximation if we include a larger number of light-object-camera configuration during training.
\item[\textbf{H3}] Increasing the number of layers in the network will result in faster convergence during training, but will negatively affect the execution time.
\item[\textbf{H4}] Generating the neural network with different ground truths will have no impact on the approximation, as long as the structure of the scene is maintained in both the input and the ground truth.
\end{enumerate}

\begin{figure}[htb]
  \centering
  \includegraphics[width=3in]{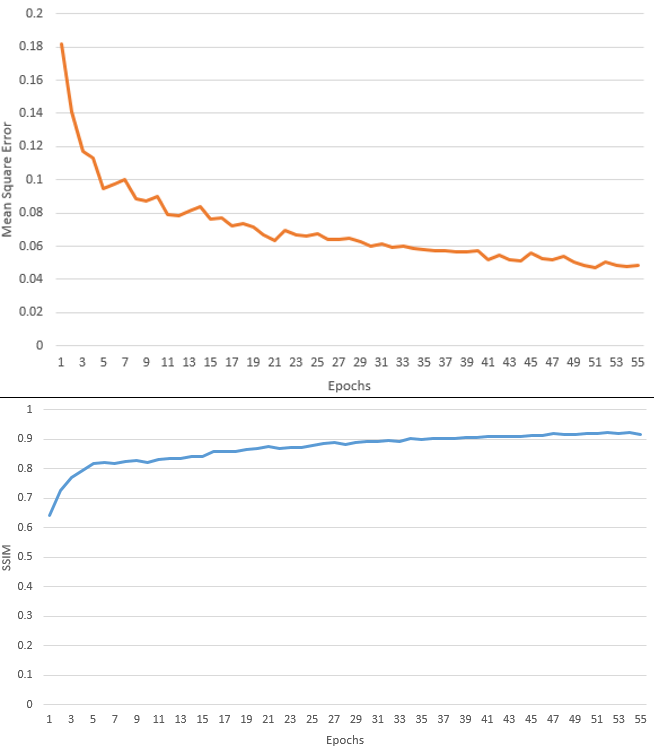}
  \caption{\label{chart:4_0_epochs.png}
   This figure shows mean square error on a test dataset after each epoch (top) and their structural similarity index (bottom).}
 
\end{figure}

\subsection{Experiment 1: Number of Training Iterations} \label{NumberIts}
We trained our network with a base layer of K=128 for 55 epochs with a training set contain 3400 image space buffers and it's corresponding ground truth. After each epoch we test the network against the intermediate frames which are unknown to the network. We measure mean square error (MSE) for per pixel error between the generated image and ground truth, and structural similarity index (SSIM) as a perceptual metric. As shown in \autoref{chart:4_0_epochs.png} (top), MSE decreases with more number of iteration which suggest that network is converging. As shown in \autoref{chart:4_0_epochs.png} (bottom), SSIM increases after each iteration which indicates that our network produces outputs which are closer to the ground truth perceptually.


\begin{figure}[htb]
  \centering
  \includegraphics[width=3.25in]{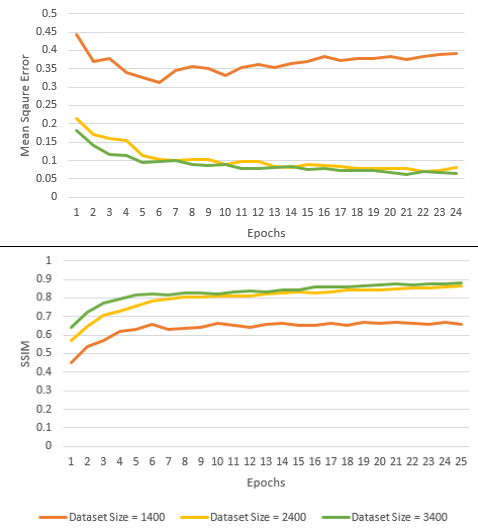}
  \caption{\label{chart:4_1_sizes}
  This figure shows a comparison of training sets with different dataset sizes. We show mean square error (top) and structure similarity index (bottom) for all training sets.}
\end{figure}

\subsection{Experiment 2: Size of Training Set} \label{SizeOfTraining}
Another experiment was conducted with different sizes (1400, 2400, and 3400) of the training set. The hyperparameters used for the networks were K=128 and 25 epochs. We found that as the training set size increases, both MSE and SSIM improve significantly and the network also converges faster, as shown in \autoref{chart:4_1_sizes}. With smaller training sets, we end up underfitting the network and end up giving a bad approximation for the unseen test data. When using a large training set, care should be given to make sure that there is enough variety in the data for the network to avoid overfitting. An optimal training dataset size depends on how big the scene is and the number of light-object-camera variations.

\begin{figure*}[htb]
  \centering
  \includegraphics[width=\linewidth]{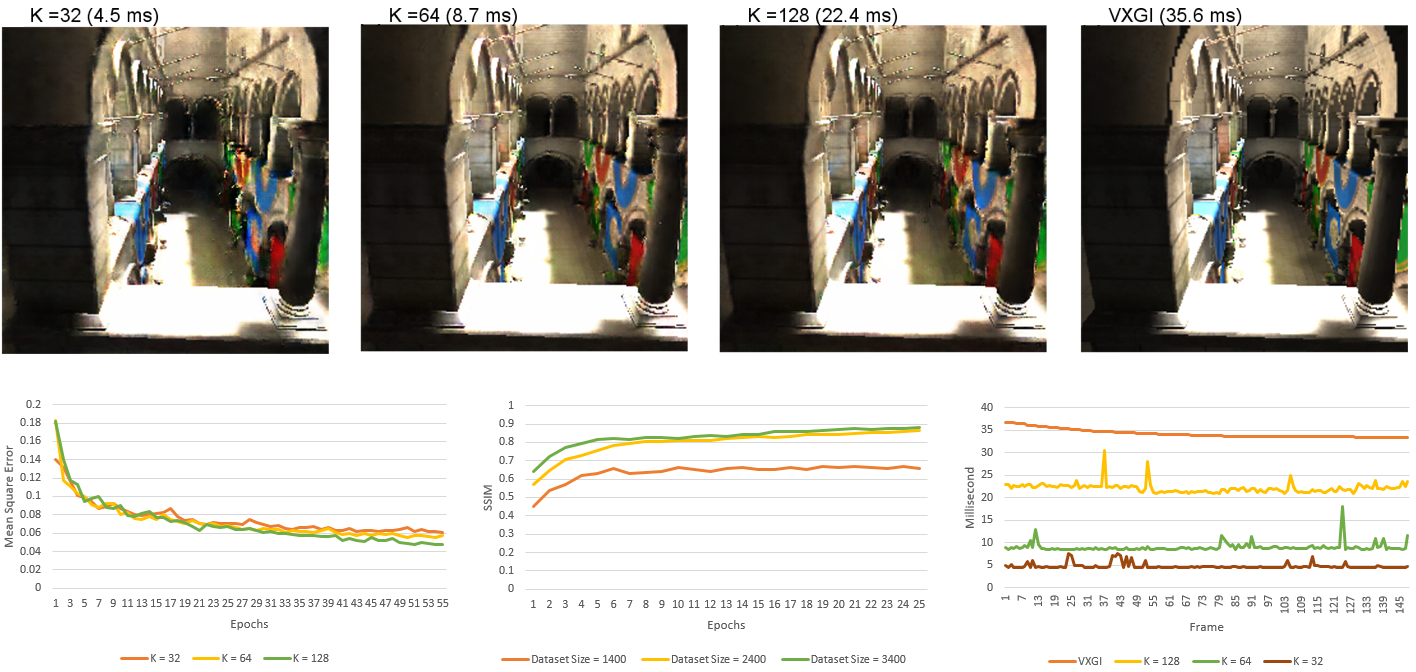}
  \caption{\label{chart:4_2_layers}
  This figure shows output from networks with different number of base layer and Voxel Cone Tracing(VXGI) (top) and comparison of the  networks in terms of mean square error (bottom left), structure similarity index (bottom middle) and runtime (bottom right).}
\end{figure*}

\subsection{Experiment 3: Number of Layers} \label{NumberOfLayers}
We found that the number of layers in the network has a significant impact on the approximation in terms of accuracy and execution speed. We used 3 separate networks with base layer K=32, 64 and 128 on a training set of 3400 images for 50 epochs. We found that a network with more number of layers converges faster but lowers the execution speed (see \autoref{chart:4_2_layers}, left and center). Our network with base layer K=32 was 10 times faster than the ground truth technique  and produces images closer to the ground truth (in this case \textit{VXGI}). As we increase the base layer to K=128 we see that even though the execution time increases, it is still faster than the ground truth technique (see \autoref{chart:4_2_layers}, right). When we map our network to an offline technique like path tracing, our network produces a similar quality image nearly 100 times faster. Example output from \textit{VXGI} vs. \textit{Deep Illumination} with the different base layers is presented in \autoref{chart:4_2_layers} (top). Note that errors would be hard to spot in a dynamic scene, even at K=32, while at K=128, differences between VXGI are nearly imperceptible, although if you zoom in on particular sections of the image, you can see, for example, that the banners hanging off the right balcony improve dramatically in quality when moving from K=32 to K=64. Likewise, the contrasts between lighter and darker areas are better approximate the ground truth in K=128 than in K=64. Although this experiment is quite promising, further exploration is needed in order to find an optimal base layer K with respect to a particular scene.

\begin{figure}[htb]
  \centering
  \includegraphics[width=3.25in]{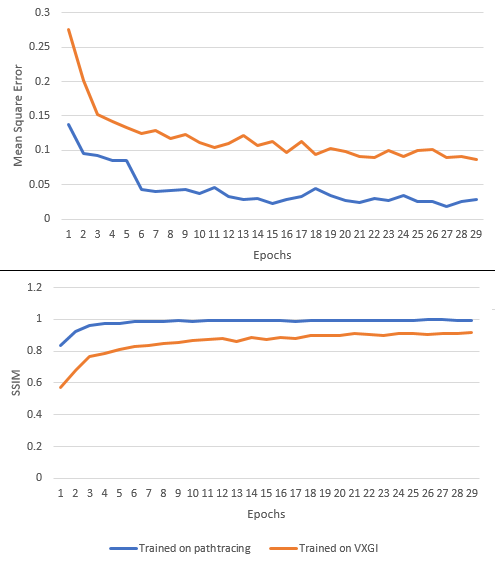}
  \caption{\label{chart:4_3_pathtracing_vs_vxgi}
  This figure shows a comparison between two different ground truths used for training (pathtracing and voxel cone tracing (VXGI) in terms of mean square error (top) and structural similarity index (bottom)}
\end{figure}

\subsection{Experiment 4: Training with Different ground truths} \label{DifferentGTs}
The same scene was rendered using a GPU based path tracing and voxel based global illumination; here we extracted 400 images for the training set. We used a base layer K=128 and trained for 30 epochs. The result shows that the network trained on path tracing converges faster than the network trained on VXGI, as indicated by the mean square error values shown in \autoref{chart:4_3_pathtracing_vs_vxgi} (top). Comparing the structural similarity (SSIM), the network trained on path tracing produces images similar to the ground truth with high accuracy under significantly less number of iteration compared to the network trained on VXGI. As shown in \autoref{chart:4_3_pathtracing_vs_vxgi} (bottom), after only 10 epochs, the SSIM of path tracing is already close to 1, whereas even after 30 epochs, VXGI is still improving.

\section{Discussion of Results} \label{Discussion}
Overall, our evaluations of different networks and dataset configurations indicates that our technique can be used in the development pipeline for game with dynamic 3D environments when we have some prior knowledge about the environment, such as its artistic style, the number of lights in the scene, the possible camera positions and orientation, and so on. As expected, Experiment 1 (Sec.~\ref{NumberIts}) shows that an increase in training iterations both reduces mean square error and  perceptual error, and so \textbf{H1} is confirmed.

As we saw in Experiment 2 (Sec.~\ref{SizeOfTraining}), an increase in the size of the training set will improve results. Additionally, more variety in the dataset helps to prevent mode collapse, which is a common problem with GANs where the network only learns only a part of the data distribution. The optimal number of epochs (or iterations) needed for a particular scene is hard to predict, but there seems to be a correlation with the complexity of the scene and number of iterations required. Using the validation set, and aided by the MSE and SSIM metrics, we can track the progress of the training and stop the training stage once we no longer observe any significant improvement in the validation set. That is, \textbf{H2} is confirmed. However, our initial exploration shows that there are diminishing returns in increasing the training set size larger than 2400. Further work needs to be done to determine the relationship between the ideal number of layers and the complexity of the scene.
 
Experiment 3 (Sec.~\ref{NumberOfLayers}) shows that the number layers have a significant impact on the execution speed. However, even though the network with base layer K=128 runs slower than the networks with base layer K=64 and K=32, compared to VXGI or path tracing it runs significantly faster. Thus \textbf{H3} is partially confirmed, as indeed the increase in the number of layers increases the running time of generating GI. However, the convergence rates are quite similar, and the running time is still faster than the ground truth technique that it is approximating.    

Experiment 4 (Sec.~\ref{DifferentGTs}) shows that learning a more accurate GI technique not only produces better results overall, but also allows the network to converge more quickly. When comparing the ground truth used for training, we see that network learns from path tracing faster than VXGI and produces structurally similar output in only 10 epochs. Thus, \textbf{H4} is \textit{not} confirmed. These results are quite unexpected and merit further investigation. We speculate that the approximation methods used in VXGI may emphasize different features in different scene configurations, and thus can introduce contradictions during the training, requiring a longer training time to learn GI features for a scene. Even though path tracing performs better and converges faster, the use of VXGI may still make sense for training purposes in some situations, during development for instance, since each ground truth example can be generated quickly, and because the perceptual quality of the GI produced is sufficient for dynamic scenes with a moving camera or moving lights and objects.

Overall, these results show that there is considerable potential for using GANs to generate high-quality GI in dynamic scenarios, and that it is practical to train a GAN to approximate GI even on a consumer GPU in a standard desktop environment, allowing \textit{Deep Illumination} to learn features specific to that scene. \autoref{figure:comparisons_and_closeups}(a) shows a result generated from a network trained using path tracing that is, to our eyes, imperceptible from the ground truth, while \mbox{\autoref{figure:comparisons_and_closeups}(b--f)} shows a range of high-quality results generated from networks trained using VXGI. Although within dynamic scenes it is hard to notice differences between the ground truth and our approximation, \autoref{figure:artifacts} demonstrates instances where \textit{Deep Illumination} incorrectly added GI to a frame. These artifacts can occur when rendering a particle system (a), and also when introducing new objects into the scene (b). However, we believe that the severity of these artifacts could be reduced or eliminated with further training that included a wider range of examples from a specific scenario. Moreover, using a more accurate GI solution (like path tracing) should also reduce these artifacts.

\section{Conclusion and Future Work}
In this paper, we presented \textit{Deep Illumination}, the first application of generative adversarial networks to the problem of global illumination, producing temporally coherent indirect illumination and soft shadows from image space buffers. There are several interesting areas to explore as future work for this approach. For example, we showed how a generative adversarial network can be used learn a mapping from G-buffer and direct illumination buffer to any global illumination solution for scenes with diffuse materials. For future exploration, we would like to study how the network behaves when introducing different materials, and we also believe that our technique will be effective at learning other global illumination effects, such as motion blur, depth of field, and caustics. 

Although our experiment shows promising results when introducing an unknown object into the scene, it will also be interesting to investigate more complex scenarios containing a wider variation of models, cameras, and lights in different positions and orientation. One interesting direction is to explore global illumination in procedurally generated environments, where the network is trained on random scenes that contain the building blocks of the environment. In terms of network structure, we plan to perform more experiments on scenes of different sizes and scales, and with different numbers of objects and lights in order to find an optimal number for the base layer, which directly impacts the runtime. 

Finally, an interesting avenue of exploration is to automate the data acquisition process for training. That is, given a particular scenario, we should be able to automatically configure the camera within possible scenes in order to extract frames for training that contain significant light-object variations. Future work will investigate optimizing the training of the network to both speed up the training process and to reduce artifacts when generating global illumination effects.

\bibliographystyle{eg-alpha-doi}

\bibliography{DI}

\newcommand{\etalchar}[1]{$^{#1}$}
\begin{thebibliography}{\uppercase{GPAM{\etalchar{*}}14}}

\bibitem[AMHH08]{Akenine-Moller:2008:RR:2829183}
\textsc{Akenine-Moller T., Haines E., Hoffman N.}:
\newblock \emph{Real-Time Rendering}, 3rd~ed.
\newblock A. K. Peters, Ltd., Natick, MA, USA, 2008.

\bibitem[Chr06]{Christensen2006}
\textsc{Christensen P.}:
\newblock {Ray Tracing for the Movie "Car"}.
\newblock \emph{2006 IEEE Symposium on Interactive Ray Tracing} (2006).
\newblock \href {http://dx.doi.org/10.1109/RT.2006.280207}
  {\path{doi:10.1109/RT.2006.280207}}.

\bibitem[CNS{\etalchar{*}}14]{Crassin2014}
\textsc{Crassin C., Neyret F., Sainz M., Green S., Eisemann E.}:
\newblock {Interactive indirect illumination using voxel cone tracing}.
\newblock \emph{Computer Graphics Forum 30}, 7 (2014), 1921--1930.
\newblock \href {http://arxiv.org/abs/1105-} {\path{arXiv:1105-}}, \href
  {http://dx.doi.org/10.1111/j.1467-8659.2011.02063.x}
  {\path{doi:10.1111/j.1467-8659.2011.02063.x}}.

\bibitem[DCF{\etalchar{*}}15]{denton2015deep}
\textsc{Denton E.~L., Chintala S., Fergus R., et~al.}:
\newblock Deep generative image models using a laplacian pyramid of adversarial
  networks.
\newblock In \emph{Advances in Neural Information Processing Systems (NIPS)}
  (2015), pp.~1486--1494.

\bibitem[EASW09]{Eisemann2009}
\textsc{Eisemann E., Assarsson U., Schwarz M., Wimmer M.}:
\newblock {Casting Shadows in Real Time}.
\newblock \emph{ACM SIGGRAPH ASIA 2009 Courses on - SIGGRAPH ASIA '09} (2009),
  21.
\newblock URL: \url{http://dl.acm.org/citation.cfm?id=1722963}, \href
  {http://dx.doi.org/10.1145/1665817.1722963}
  {\path{doi:10.1145/1665817.1722963}}.

\bibitem[ED10]{engelhardt2010epipolar}
\textsc{Engelhardt T., Dachsbacher C.}:
\newblock Epipolar sampling for shadows and crepuscular rays in participating
  media with single scattering.
\newblock In \emph{Proceedings of the 2010 ACM SIGGRAPH symposium on
  Interactive 3D Graphics and Games} (2010), ACM, pp.~119--125.

\bibitem[Gau14]{Gauthier2014}
\textsc{Gauthier J.}:
\newblock {Conditional generative adversarial nets for convolutional face
  generation}.
\newblock \emph{Class Project for Stanford CS231N: Convolutional Neural
  Networks for Visual Recognition, Winter semester 2014} (2014).
\newblock URL:
  \url{http://cs231n.stanford.edu/reports/jgauthie{\_}final{\_}report.pdf},
  \href {http://arxiv.org/abs/1512.09300} {\path{arXiv:1512.09300}}.

\bibitem[GEB16]{Gatys2016}
\textsc{Gatys L.~A., Ecker A.~S., Bethge M.}:
\newblock {Image style transfer using convolutional neural networks}.
\newblock \emph{The IEEE conference on computer vision and pattern recognition}
  (2016), 2414--2423.
\newblock \href {http://arxiv.org/abs/1505.07376} {\path{arXiv:1505.07376}},
  \href {http://dx.doi.org/10.1109/CVPR.2016.265}
  {\path{doi:10.1109/CVPR.2016.265}}.

\bibitem[GPAM{\etalchar{*}}14]{Goodfellow2014}
\textsc{Goodfellow I., Pouget-Abadie J., Mirza M., Xu B., Warde-Farley D.,
  Ozair S., Courville A., Bengio Y.}:
\newblock {Generative Adversarial Nets}.
\newblock \emph{Advances in Neural Information Processing Systems 27} (2014),
  2672--2680.
\newblock URL:
  \url{http://papers.nips.cc/paper/5423-generative-adversarial-nets.pdf}, \href
  {http://arxiv.org/abs/arXiv:1406.2661v1} {\path{arXiv:arXiv:1406.2661v1}},
  \href {http://dx.doi.org/10.1017/CBO9781139058452}
  {\path{doi:10.1017/CBO9781139058452}}.

\bibitem[Hou06]{Houska2006}
\textsc{Houska P.}:
\newblock {Directional Lightmaps}.
\newblock \emph{Image (Rochester, N.Y.)} (2006).

\bibitem[HSK16]{Holden2016}
\textsc{Holden D., Saito J., Komura T.}:
\newblock {Neural network ambient occlusion}.
\newblock In \emph{SIGGRAPH ASIA 2016 Technical Briefs on - SA '16} (2016),
  pp.~1--4.
\newblock URL: \url{http://dl.acm.org/citation.cfm?doid=3005358.3005387}, \href
  {http://dx.doi.org/10.1145/3005358.3005387}
  {\path{doi:10.1145/3005358.3005387}}.

\bibitem[HZRS16]{He2016}
\textsc{He K., Zhang X., Ren S., Sun J.}:
\newblock {Identity Mappings in Deep Residual Networks Importance of Identity
  Skip Connections Usage of Activation Function Analysis of Pre-activation
  Structure}.
\newblock In \emph{ECCV 2016} (2016), no.~1, pp.~1--15.
\newblock URL: \url{https://github.com/KaimingHe/resnet-1k-layers}, \href
  {http://arxiv.org/abs/1603.05027} {\path{arXiv:1603.05027}}, \href
  {http://dx.doi.org/10.1007/978-3-319-46493-0_38}
  {\path{doi:10.1007/978-3-319-46493-0_38}}.

\bibitem[IZZE16]{Isola2016}
\textsc{Isola P., Zhu J.-Y., Zhou T., Efros A.~A.}:
\newblock {Image-to-Image Translation with Conditional Adversarial Networks}.
\newblock URL: \url{http://arxiv.org/abs/1611.07004}, \href
  {http://arxiv.org/abs/1611.07004} {\path{arXiv:1611.07004}}.

\bibitem[KD10]{Kaplanyan2010}
\textsc{Kaplanyan A., Dachsbacher C.}:
\newblock {Cascaded light propagation volumes for real-time indirect
  illumination}.
\newblock \emph{on Interactive 3D Graphics and Games 1}, 212 (2010), 99.
\newblock URL:
  \url{http://portal.acm.org/citation.cfm?doid=1730804.1730821{\%}5Cnhttp://dl.acm.org/citation.cfm?id=1730821},
  \href {http://dx.doi.org/10.1145/1730804.1730821}
  {\path{doi:10.1145/1730804.1730821}}.

\bibitem[LTH{\etalchar{*}}16]{DBLP:journals/corr/LedigTHCATTWS16}
\textsc{Ledig C., Theis L., Huszar F., Caballero J., Aitken A.~P., Tejani A.,
  Totz J., Wang Z., Shi W.}:
\newblock Photo-realistic single image super-resolution using a generative
  adversarial network.
\newblock \emph{CoRR abs/1609.04802} (2016).
\newblock URL: \url{http://arxiv.org/abs/1609.04802}.

\bibitem[LW93]{Lafortune1993}
\textsc{Lafortune E.~P., Willems Y.~D.}:
\newblock {Bi-Directional Path Tracing}.
\newblock \emph{Proc. SIGGRAPH} (1993), 145--153.
\newblock URL:
  \url{http://citeseerx.ist.psu.edu/viewdoc/summary?doi=10.1.1.28.3913}, \href
  {http://arxiv.org/abs/arXiv:1011.1669v3} {\path{arXiv:arXiv:1011.1669v3}},
  \href {http://dx.doi.org/10.1017/CBO9781107415324.004}
  {\path{doi:10.1017/CBO9781107415324.004}}.

\bibitem[MHN13]{Maas2013}
\textsc{Maas A.~L., Hannun A.~Y., Ng A.~Y.}:
\newblock {Rectifier Nonlinearities Improve Neural Network Acoustic Models}.
\newblock \emph{Proceedings of the 30 th International Conference on Machine
  Learning 28} (2013), 6.
\newblock URL:
  \url{https://web.stanford.edu/{~}awni/papers/relu{\_}hybrid{\_}icml2013{\_}final.pdf}.

\bibitem[MMNL16]{Mara2016}
\textsc{Mara M., McGuire M., Nowrouzezahrai D., Luebke D.}:
\newblock \emph{{Deep G-Buffers for Stable Global Illumination Approximation}}.
\newblock Tech. rep., 2016.
\newblock \href {http://dx.doi.org/10.2312/hpg.20161195}
  {\path{doi:10.2312/hpg.20161195}}.

\bibitem[MO14]{DBLP:journals/corr/MirzaO14}
\textsc{Mirza M., Osindero S.}:
\newblock Conditional generative adversarial nets.
\newblock \emph{CoRR abs/1411.1784} (2014).
\newblock URL: \url{http://arxiv.org/abs/1411.1784}.

\bibitem[NAM{\etalchar{*}}17]{Nalbach2017}
\textsc{Nalbach O., Arabadzhiyska E., Mehta D., Seidel H.~P., Ritschel T.}:
\newblock {Deep Shading: Convolutional Neural Networks for Screen Space
  Shading}.
\newblock \emph{Computer Graphics Forum 36}, 4 (2017), 65--78.
\newblock \href {http://dx.doi.org/10.1111/cgf.13225}
  {\path{doi:10.1111/cgf.13225}}.

\bibitem[NW10]{Nichols2010}
\textsc{Nichols G., Wyman C.}:
\newblock {Interactive indirect illumination using adaptive multiresolution
  splatting}.
\newblock \emph{IEEE Transactions on Visualization and Computer Graphics 16}, 5
  (2010), 729--741.
\newblock \href {http://dx.doi.org/10.1109/TVCG.2009.97}
  {\path{doi:10.1109/TVCG.2009.97}}.

\bibitem[RDGK12]{Ritschel2012}
\textsc{Ritschel T., Dachsbacher C., Grosch T., Kautz J.}:
\newblock {The state of the art in interactive global illumination}.
\newblock \emph{Computer Graphics Forum 31}, 1 (2012), 160--188.
\newblock \href {http://dx.doi.org/10.1111/j.1467-8659.2012.02093.x}
  {\path{doi:10.1111/j.1467-8659.2012.02093.x}}.

\bibitem[RFB15]{Ronneberger2015}
\textsc{Ronneberger O., Fischer P., Brox T.}:
\newblock {U-Net: Convolutional Networks for Biomedical Image Segmentation}.
\newblock \emph{Miccai} (2015), 234--241.
\newblock \href {http://arxiv.org/abs/1505.04597} {\path{arXiv:1505.04597}},
  \href {http://dx.doi.org/10.1007/978-3-319-24574-4_28}
  {\path{doi:10.1007/978-3-319-24574-4_28}}.

\bibitem[RGS09]{Ritschel2009}
\textsc{Ritschel T., Grosch T., Seidel H. H.-P.}:
\newblock {Approximating Dynamic Global Illumination in Image Space}.
\newblock In \emph{Proceedings of the 2009 symposium on Interactive 3D graphics
  and games - I3D '09} (2009), pp.~75--82.
\newblock URL:
  \url{http://portal.acm.org/citation.cfm?doid=1507149.1507161{\%}5Cnhttp://portal.acm.org/citation.cfm?id=1507161},
  \href {http://dx.doi.org/10.1145/1507149.1507161}
  {\path{doi:10.1145/1507149.1507161}}.

\bibitem[RGW{\etalchar{*}}06]{Ren2006}
\textsc{Ren Z., Guo B., Wang R., Snyder J., Zhou K., Liu X., Sun B., Sloan
  P.-P., Bao H., Peng Q.}:
\newblock {Real-time soft shadows in dynamic scenes using spherical harmonic
  exponentiation}.
\newblock \emph{ACM Transactions on Graphics 25}, 3 (2006), 977.
\newblock URL: \url{http://dl.acm.org/citation.cfm?id=1141982}, \href
  {http://dx.doi.org/10.1145/1141911.1141982}
  {\path{doi:10.1145/1141911.1141982}}.

\bibitem[SHR10]{Soler2010}
\textsc{Soler C., Hoel O., Rochet F.}:
\newblock {A deferred shading pipeline for real-time indirect illumination}.
\newblock \emph{ACM SIGGRAPH 2010 Talks 5} (2010), 2009.
\newblock URL: \url{http://dl.acm.org/citation.cfm?id=1837049}, \href
  {http://dx.doi.org/10.1145/1837026.1837049}
  {\path{doi:10.1145/1837026.1837049}}.

\bibitem[SKP07]{Shah2007}
\textsc{Shah M.~A., Konttinen J., Pattanaik S.}:
\newblock {Caustics mapping: An image-space technique for real-time caustics}.
\newblock \emph{IEEE Transactions on Visualization and Computer Graphics 13}, 2
  (2007), 272--280.
\newblock \href {http://dx.doi.org/10.1109/TVCG.2007.32}
  {\path{doi:10.1109/TVCG.2007.32}}.

\bibitem[SKS02]{Sloan2002}
\textsc{Sloan P.-P., Kautz J., Snyder J.}:
\newblock {Precomputed radiance transfer for real-time rendering in dynamic,
  low-frequency lighting environments}.
\newblock \emph{ACM Transactions on Graphics 21}, 3 (2002).
\newblock URL: \url{http://portal.acm.org/citation.cfm?doid=566654.566612},
  \href {http://dx.doi.org/10.1145/566654.566612}
  {\path{doi:10.1145/566654.566612}}.

\bibitem[SSMW09]{Scherzer2009}
\textsc{Scherzer D., Schw??rzler M., Mattausch O., Wimmer M.}:
\newblock {Real-time soft shadows using temporal coherence}.
\newblock In \emph{Lecture Notes in Computer Science (including subseries
  Lecture Notes in Artificial Intelligence and Lecture Notes in
  Bioinformatics)} (2009), vol.~5876 LNCS, pp.~13--24.
\newblock \href {http://dx.doi.org/10.1007/978-3-642-10520-3_2}
  {\path{doi:10.1007/978-3-642-10520-3_2}}.

\bibitem[ST90]{Saito1990}
\textsc{Saito T., Takahashi T.}:
\newblock {Comprehensible rendering of 3-D shapes}.
\newblock \emph{ACM SIGGRAPH Computer Graphics 24}, 4 (1990), 197--206.
\newblock URL: \url{http://portal.acm.org/citation.cfm?doid=97880.97901}, \href
  {http://dx.doi.org/10.1145/97880.97901} {\path{doi:10.1145/97880.97901}}.

\bibitem[TO12]{Tokuyoshi2012}
\textsc{Tokuyoshi Y., Ogaki S.}:
\newblock {Real-Time Bidirectional Path Tracing via Rasterization}.
\newblock In \emph{Proceedings of the ACM SIGGRAPH Symposium on High
  Performance Graphics} (2012), pp.~183--190.
\newblock URL: \url{http://dl.acm.org/citation.cfm?id=2159647}, \href
  {http://dx.doi.org/10.1145/2159616.2159647}
  {\path{doi:10.1145/2159616.2159647}}.

\bibitem[WKBK02]{Wald2002}
\textsc{Wald I., Kollig T., Benthin C., Keller A.}:
\newblock {Interactive Global Illumination using Fast Ray Tracing}.
\newblock \emph{Eurographics Workshop on Rendering} (2002), 11.

\bibitem[YYZ{\etalchar{*}}17]{DBLP:journals/corr/abs-1708-00961}
\textsc{Yang Q., Yan P., Zhang Y., Yu H., Shi Y., Mou X., Kalra M.~K., Wang
  G.}:
\newblock Low dose {CT} image denoising using a generative adversarial network
  with wasserstein distance and perceptual loss.
\newblock \emph{CoRR abs/1708.00961} (2017).
\newblock URL: \url{http://arxiv.org/abs/1708.00961}.

\bibitem[ZJL17]{DBLP:journals/corr/ZhangJL17}
\textsc{Zhang L., Ji Y., Lin X.}:
\newblock Style transfer for anime sketches with enhanced residual u-net and
  auxiliary classifier {GAN}.
\newblock \emph{CoRR abs/1706.03319} (2017).
\newblock URL: \url{http://arxiv.org/abs/1706.03319}.

\end{thebibliography}

\end{document}